\documentclass[a4paper]{ISOStyle}
\usepackage{epsfig}

\newcommand{\cc}{$\mathrm{^{12}C/^{13}C}$}
\newcommand{\ad}{$\theta_d$}
\newcommand{\vt}{$\xi_t$}
\newcommand{\kms}{$\rm{km\,s^{-1}}$}
\newcommand{\teff}{T$_{\mathrm{eff}}$}
\newcommand{\mic}{$\mu$m}

\newcommand{\Msun}{M$_{\odot}$}

\begin{document}

\title{Stellar models in IR calibration}

\author{L.\,Decin}
 \institute{Instituut voor Sterrenkunde, KULeuven, Celestijnenlaan 200B, B-3001
    Leuven, Belgium}

\maketitle 

\begin{abstract}

For the astronomical community analyzing ISO-SWS data, a first point
to assess when judging and qualifying the observational data concerns
the flux calibration accuracy. Since the calibration process is not
straightforward and since a wrong calibration may lead to an over- or
underestimation of the results, knowledge on the full calibration
process and on the still remaining calibration problems is crucial
when processing the data.

One way to detect calibration problems is by comparing observed data
with theoretical predictions of a whole sample of calibration
sources. By using an iterative process in which improvements on both
the calibration and the theoretical modelling are involved, we will
demonstrate that a consistent theoretical data-set of infrared
spectra has been constructed. This data-set has been used to derive 
the OLP10 flux calibration of the ISO-SWS. We will shown
that the relative flux calibration accuracy of (high-flux) ISO-SWS
observations has reached a 2\,\% level of accuracy in band 1 and that
the flux calibration accuracy in band 2 has improved significantly
with the introduction of the memory-effect corrrection and the use of
our synthetic spectra for the flux calibration derivation, reaching a
6\,\% level of accuracy in band 2. 

\end{abstract}

\vspace*{-4ex}
\section{INTRODUCTION: SCIENTIFIC GOALS}\label{LD_sec:scgoals}

The modelling and interpretation of the ISO-SWS data requires an
accurate calibration of the spectrometers
(\cite{LD_Schipman2001}). For this purpose, many calibration sources
are observed. In the SWS spectral region (2.38 -- 45.2\,$\mu$m) the
primary standard calibration candles are bright, mostly cool,
stars. The better the behaviour of these calibration sources in the
infrared is known, the more accurate the spectrometers will be
calibrated. ISO however offered the astronomical community the first
opportunity to perform spectroscopic observations between 2.38 and
45.2\,$\mu$m at a spectral resolving power of $\sim 1500$, not polluted
by any molecular absorption caused by the earth's
atmosphere. Consequently the theoretical predictions for these sources
are not perfect due to errors in the computation of the stellar models
or in the generation of the synthetic spectra or due to the
restriction of our knowledge on the mid-infrared behaviour of these
sources.  A full exploitation of the ISO-SWS data can therefore only
result from an {\it{iterative}} process in which both new theoretical
developments on the computation of the stellar spectra and more
accurate instrumental calibration are involved. Till now, the detailed
spectroscopic analysis of the ISO-SWS data in the framework of this
iterative process has been restricted to the wavelength
region from 2.38 to 12\,$\mu$m. So, if not specified, the wavelength
range under research is limited to band 1 (2.38 -- 4.08\,$\mu$m) and
band 2 (4.08 -- 12.00\,$\mu$m).

This paper is organized as follows: in Sect.\ \ref{LD_sec:method} the
method of analysis will be shortly described. An overview of the main
results will be given in Sect.\ \ref{LD_sec:results}, while the impact
on the calibration of the SWS spectrometers will be elaborated
on in Sect.\ \ref{LD_sec:impact}. The resultant set of 16 IR synthetic
spectra will be compared with the SEDs by Cohen (\cite{LD_cohen1992},
\cite{LD_cohen1995}, 
\cite{LD_cohen1996}, \cite{LD_witteborn1999}) in Sect.\
\ref{LD_sec:cohen} and in the last section, Sect.\ \ref{LD_sec:concl}
we will end with some lessons learned from ISO for future instruments.

\vspace*{-2ex}
\section{METHOD OF ANALYSIS} \label{LD_sec:method}

Precisely because this research involves an iterative process, one has
to be extremely careful not to confuse technical detector problems
with astrophysical issues. Therefore several precautions are
taken and the analysis in its entirety enclosed several steps
including 1.\ a spectral cover of standard infrared sources from A0 to
M8, 2.\ a homogeneous way of data reduction, 3.\ a detailed literature
study, 4.\ a detailed knowledge of the impact of the various stellar
parameters on the spectral signature, 5.\ a statistical method to test
the goodness-of-fit (Kolmogorov-Smirnov test) and
6.\ high-resolution observations with two independent instruments.   
All of these steps are elaborated on in \cite*{LD_decin2000} and
\cite*{LD_decin2001_II}. 

\vspace*{-2ex}
\section{RESULTS} \label{LD_sec:results}

Using this method, all {\it{hot}} stars in our sample (i.\ e.\ stars
whose effective 
temperature is higher than the effective temperature of the Sun) and
{\it{cool}} stars are analyzed carefully. Computing for the first time
synthetic spectra in the infrared
by using the MARCS-code (\cite*{LD_gustafsson1975} and subsequent
updates) is one step, distilling useful information from it is a
second --- and far more difficult --- one. Fundamental stellar
parameters for the sample of calibration sources are a first direct
result. In papers III and IV in the series `ISO-SWS calibration and
the accurate modelling of cool-star atmospheres'
(\cite{LD_decin2001_III} and \cite{LD_decin2001_IV}) these
parameters are discussed and confronted with other published stellar parameters.

\vspace*{-3ex}
\begin{figure}[h!]
 \begin{center}
\resizebox{.45\textwidth}{!}{\rotatebox{90}{\includegraphics{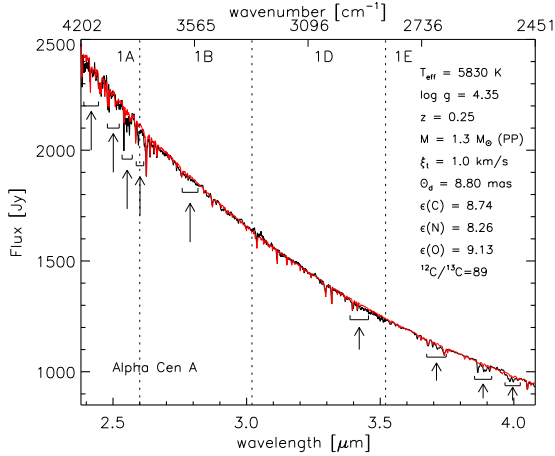}}}
  \end{center}\vspace*{-5ex}
\caption{\label{LD_fig1}Comparison between the ISO-SWS data of
$\alpha$ Cen A (black) and the synthetic spectrum (red) with
stellar parameters \teff\ = 5830\,K, $\log$ g = 4.35, M = 1.3\,\Msun, z = 0.25,
\vt\ = 1.0\,\kms, \cc\ = 89, $\varepsilon$(C) = 
8.74, $\varepsilon$(N) = 8.26, $\varepsilon$(O) = 9.13 and \ad\ =
8.80\,mas. Some of the most prominent discrepancies between these
two spectra are indicated by an arrow.}
\end{figure}
\vspace*{-1.22cm}
\begin{figure}[h!]
  \begin{center}
\resizebox{.45\textwidth}{!}{\rotatebox{90}{\includegraphics{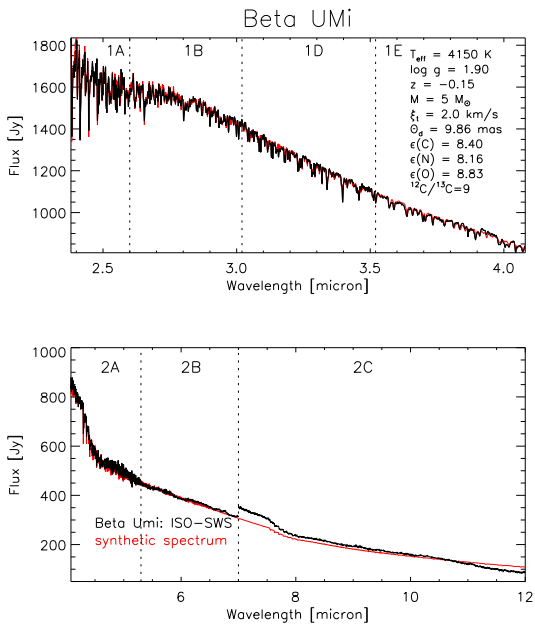}}}
  \end{center}\vspace*{-5ex}
\caption{\label{LD_fig2}Comparison between band 1 and band 2 of the
ISO-SWS data of $\beta$ UMi (black) and the synthetic spectrum
(red) with stellar parameters \teff\ = 4150\,K, $\log$ g = 1.90, M
= 5\,\Msun, z = $-0.15$, \vt\ = 2.0\,\kms, \cc\ = 9, $\varepsilon$(C) =
8.40, $\varepsilon$(N) = 8.16, $\varepsilon$(O) = 8.83 and \ad\ =
9.86\,mas.}
\end{figure}

\vspace*{-1ex}
Far more interesting are however the discrepancies which emerge from
the confrontation between ISO-SWS (OLP8.4) and theoretical spectra. A typical
example of both a {\it{hot}} and {\it{cool}} star is depicted in
Fig.\ \ref{LD_fig1} and Fig.\ \ref{LD_fig2} respectively. By
scrutinizing carefully the various discrepancies between the ISO-SWS
data and the synthetic spectra of the standard stars in our sample,
the origin of the different discrepancies was elucidated. A
description of the general trends in discrepancies for both {\it{hot}}
and {\it{cool}} stars can be found in \cite*{LD_decin2001_II}, while
the {\it{hot}} stars are discussed individually in
\cite*{LD_decin2001_III} and the {\it{cool}} stars in
\cite*{LD_decin2001_IV}. A summary of the different kinds of
discrepancies and the reason why they arise is given in Table
\ref{LD_tab:table1}.

\vspace*{-3ex}
\begin{table}[bht]
  \caption{Summary of the origins of the different types of
  discrepancies found between the ISO-SWS data and the synthetic spectra for
  the standard calibration sources in our sample. }
  \label{LD_tab:table1}
  \begin{center}
    \leavevmode
    \scriptsize
    \begin{tabular}[h]{p{.22\textwidth}|p{.22\textwidth}}
      \hline \\[-5pt]
      \multicolumn{1}{c|}{{\it{HOT}} stars } &
  \multicolumn{1}{c}{{\it{COOL}} stars}\\[+5pt]
      \hline \\[-5pt]
      1.\ hydrogen lines $\rightarrow$ {\it{problems with computation
  of hydrogenic line broadening}} & 1.\ CO $\Delta v = 2$ lines:
  predicted as too strong $\rightarrow$ {\it{inaccurate knowledge of
  resolution and instrumental profile +  problematic
  temperature distribution in outermost layers of models }} \\
     2.\ atomic features $\rightarrow$ {\it{inaccurate atomic
  oscillator strengths in IR }} & 
 2.\ atomic features $\rightarrow$ {\it{inaccurate atomic oscillator
  strenghts in IR}}\\
     3.\ band 1B -- 1D $\rightarrow$ {\it{Humphreys ionization edge}}
  & 3.\ OH-lines: predicted as too weak $\rightarrow$ {\it{problematic
  temperature distribution in outermost layers of models}} \\
     4.\ 2.38 -- 2.4\,$\mu$m $\rightarrow$ {\it{inaccurate RSRF}} &
     4.\ 2.38 -- 2.4\,$\mu$m $\rightarrow$ {\it{inaccurate RSRF}} \\
     5.\ from 3.84\,$\mu$m on $\rightarrow$ {\it{fringes}} &
     5.\ from 3.84\,$\mu$m on $\rightarrow$ {\it{fringes}}\\
     6.\ band 2 $\rightarrow$ {\it{memory-effects + inaccurate RSRF}} &
     6.\ band 2 $\rightarrow$ {\it{memory-effects + inaccurate RSRF}}\\[+5pt]
      \hline 
     \end{tabular}
  \end{center}
\end{table}

\vspace*{-5ex}
\section{IMPACT ON CALIBRATION (AND MODELLING)} \label{LD_sec:impact}

The results of the detailed comparison between observed ISO-SWS data
and synthetic spectra have their implications both on the calibration of the
ISO-SWS data and on the theoretical description of the stellar
atmospheres. Since we are here discussing the `Calibration Legacy of
the ISO Mission', the emphasis will be mainly on the calibration
issue.

\begin{figure}[h!]
\begin{center}
\resizebox{0.4\textwidth}{!}{\rotatebox{90}{\includegraphics{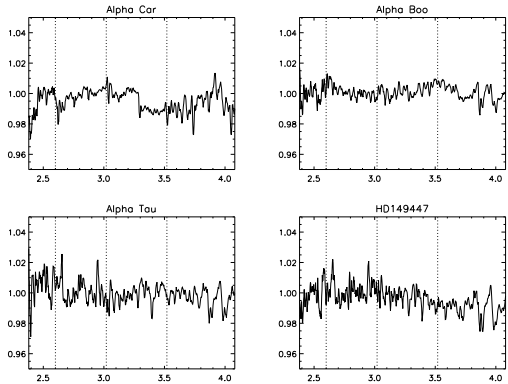}}}
\caption{\label{LD_fig3}The observed ISO-SWS spectra --- processed
with the flux calibration as in OLP8.4 --- of 4
stars with different spectral types are divided by their synthetic
spectrum. This ratio is then rebinned to a resolution of 250. Only
at the beginning of band 1A a same trend is visible for all stars,
indicating that the broad-band shape of the RSRFs is already quite
accurately known for 
the different sub-bands of band 1, with the only exception being
at the beginning of band 1A. The feature arising around 3.85\,\mic\
is an atomic feature mispredicted due to inaccurate atomic oscillatorstrengths.}
\end{center}
\end{figure}

From the calibration point of view, a first conclusion is
reached that the broad-band shape of the relative spectral response
function (RSRF) is
at the moment already quite accurate in band 1, although some improvements
can be made at the beginning of band 1A and in band 2  (see Fig.\
\ref{LD_fig3}). Also, a fringe pattern is recognized at the 
end of band 1D. The limited accuracy of our approximation of the
instrumental profile may cause the strongest CO lines to be predicted
as too strong.  
The main consequence for the further (OLP10) calibration of ISO-SWS can
however be seen from Fig.\ \ref{LD_fig3}: from this figure, it is
clearly visible that an agreement of better than 2\,\% is reached in
band 1. Since the same molecules are absorbing in band 1 and in band
2, these synthetic spectra are supposed to be also accurate in band 2. 
I.\ e. a {\it consistent} data-set of {\it 16} IR (synthetic) spectra for
stars with spectral type going from A0 till M2 has been constructed in
the wavelength range from 2.38 till 12\,$\mu$m! These
spectra were then used to test the recently developed
method for memory-effect correction (\cite{LD_Kester2001}) and to
rederive the relative spectral response function for bands 1 and 2 for
OLP10. In conjunction with photometric data (e.g.\ of M.\ Cohen) this
{\underline{same}} input data-set was then used for the re-calibration
of the absolute flux-levels of the spectra observed with ISO-SWS. In
this way, both consistency and integrity were implemented.

Moreover, this set of synthetic spectra is not only used to 
improve the flux calibration of the observations taken during the
nominal phase, but they are also an excellent tool to characterize
instabilities of the SWS spectrometers during the post-helium
mission.

\vspace*{-3ex}

\begin{figure}[h!]
  \begin{center}
\resizebox{.45\textwidth}{!}{\rotatebox{90}{\includegraphics{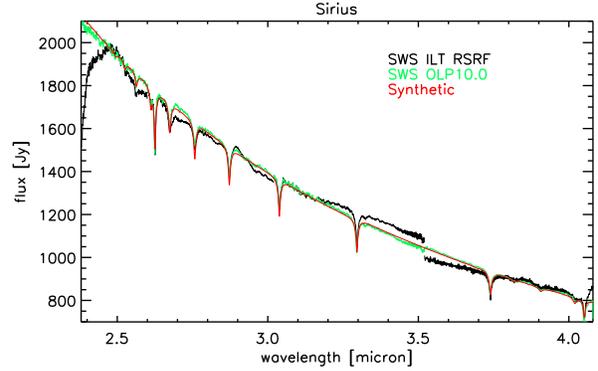}}}
  \end{center}\vspace*{-2ex}
\caption{\label{LD_fig4}Comparison between the band-1 AOT01 speed-4
observation of Sirius 
reduced by using the old ILT RSRF's (black) or the new OLP10 (green) and
the synthetic spectrum with stellar parameters \teff\ = 10150\,K, $\log$ g
= 4.30, M = 2.0\,\Msun, z = 0.50, \vt\ = 2.0\,\kms, \cc\ = 89,
$\varepsilon$(C) = 7.97, $\varepsilon$(N) = 8.15, $\varepsilon$(O)
= 8.55 and \ad\ = 6.17\,mas.}
\end{figure}

\begin{figure}[h!]
  \begin{center}
\resizebox{.45\textwidth}{!}{\rotatebox{90}{\includegraphics{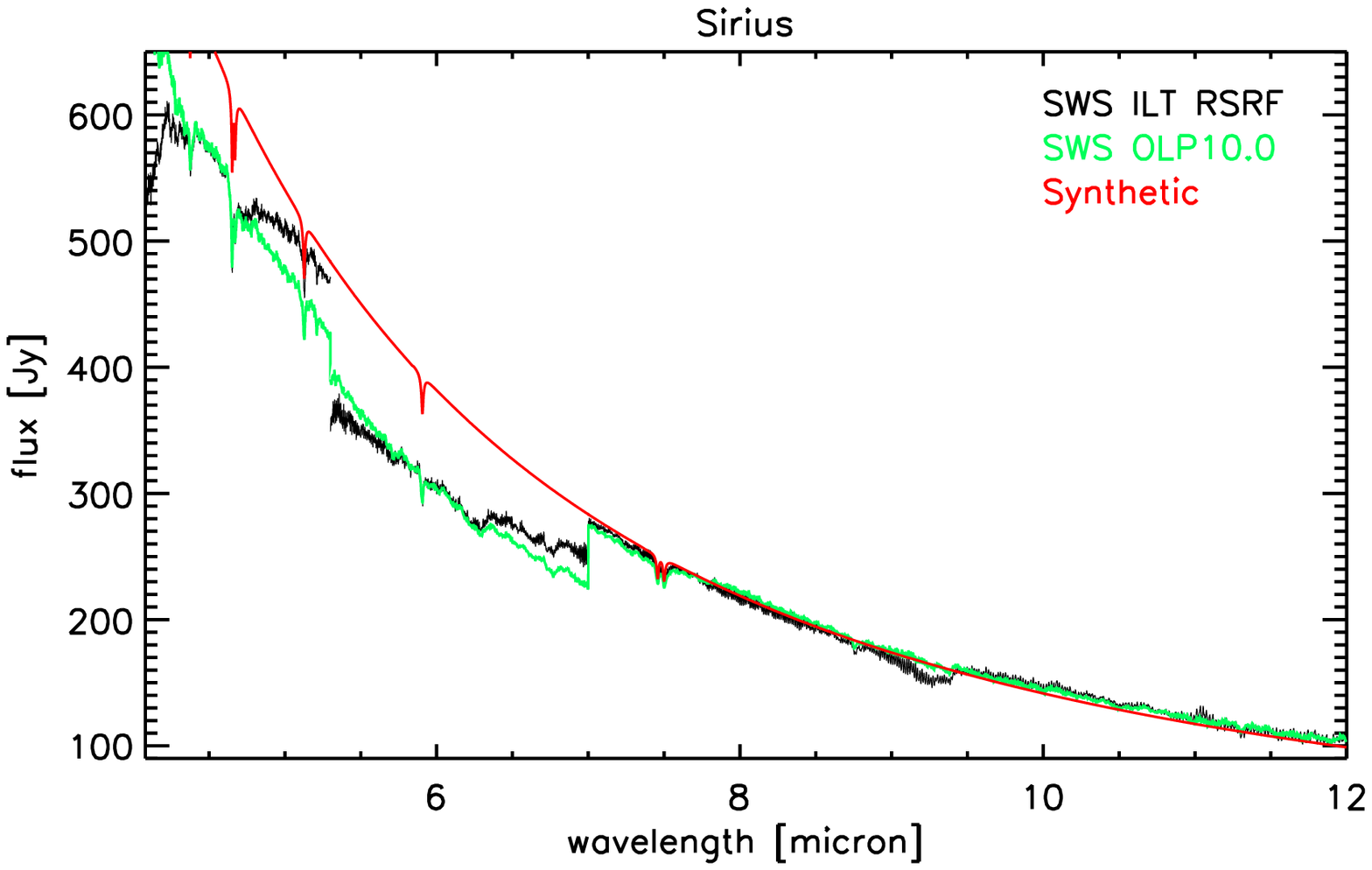}}}
  \end{center}\vspace*{-2ex}
\caption{\label{LD_fig5}Comparison between the band-2 AOT01 speed-4
observation of Sirius 
reduced by using the old ILT RSRF's (black) or the new OLP10 (green) and
the synthetic spectrum with stellar parameters \teff\ = 10150\,K, $\log$ g
= 4.30, M = 2.0\,\Msun, z = 0.50, \vt\ = 2.0\,\kms, \cc\ = 89,
$\varepsilon$(C) = 7.97, $\varepsilon$(N) = 8.15, $\varepsilon$(O)
= 8.55 and \ad\ = 6.17\,mas.}
\end{figure}

\vspace*{-4ex}
After summarizing these various calibration problems, it is instructive
to compare SWS observations reduced by using an `older' OLP version
and the in-test-phase OLP10 reduction tools. The conclusions are
illustrated with  the AOT01 speed-4 observation of Sirius
observed in revolution 689 (see Fig.\ \ref{LD_fig4} and Fig.\
\ref{LD_fig5}). During the reduction of the data, none of the sub-bands
has been shifted or tilted. 

In band 1, the situation is quite clear: Fig.\ \ref{LD_fig4} proves
that the reduction of ISO-SWS (high-flux) sources has improved
seriously during (and after) the ISO-mission. A relative accuracy of
better than 2\,\% is reached in band 1. The discrepancies ($> 2\,\%$)
still left, e.g.\ between 3.28 and 3.52\,$\mu$m, can be explained by 
shortcomings of the theoretical predictions. 

\begin{figure}[h!]
\mbox{} \\ \mbox{}\\ \mbox{}\\
\resizebox{.4\textwidth}{!}{\rotatebox{90}{\includegraphics{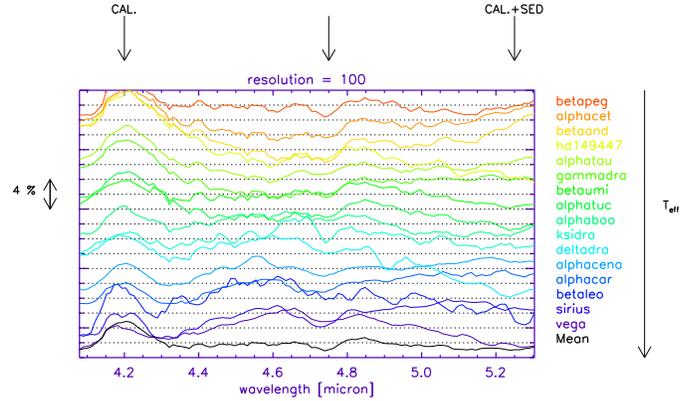}}}
\caption{\label{LD_fig6} Division between SWS AOT01 data and their theoretical
predictions at a resolving power of 100. The synthetic spectra are
calculated using as stellar parameters the values determined from the
band-1 data. In this figure, the stars are ordered by spectral
type. The mean of the different, coloured, divided spectra is given in
black at the bottom of the figure.}
\end{figure}

\begin{figure}[h!]
\mbox{} \\ \mbox{}\\ \mbox{}\\
\resizebox{.4\textwidth}{!}{\rotatebox{90}{\includegraphics{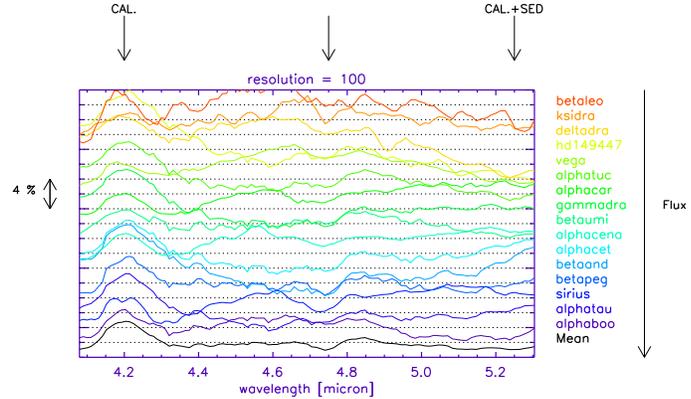}}}
\caption{\label{LD_fig7}Division between SWS AOT01 data and their theoretical
predictions at a resolving power of 100. The synthetic spectra are
calculated using as stellar parameters the values determined from the
band-1 data. In this figure, the stars are ordered by absolute-flux value.
The mean of the different, coloured, divided spectra is given in
black at the bottom of the figure.}
\end{figure}

In band 2, we can see that the calibration has improved
w.\ r.\ t.\ previous calibrations, but that still some problems do
exist. Due to the new model for memory-effect correction --- based on the
models of Fouks-Schubert (\cite{LD_Fouks2001}) and developed by Kester
(\cite{LD_Kester2001}) --- the problems with the memory-effects are
now more or less under control, at least when the flux-jump is not too
high. Consequently, the RSRFs have changed and improved a lot. The
absolute-flux calibration do, however, still predict from time to time
a wrong absolute-flux level, especially in bands 2A and 2B, the
latter one giving almost consistently across our sample too low a flux
level. When 
concentrating on the relative flux accuracy, one needs to inspect the
kind of plots as given for band 2A in Fig.\ \ref{LD_fig6} and Fig.\
\ref{LD_fig7}. The difference between the two figures is situated in
the fact that stars are ordered by spectral type in Fig. \ref{LD_fig6}
and by absolute-flux value in Fig. \ref{LD_fig7}. This is necessary to
be able to distinguish between problems which are spectral type
related --- i.\ e.\ problems with the theoretical predictions --- and
problems which are flux, and so memory-effect, related. From 
an inspection of these figures we can learn that:
\begin{itemize}
\item{One can clearly see a bump of $\sim 2.5$\,\% in the
wavelength-range from 4.08 -- 4.30\,$\mu$m. The presence of this bump
in hot as well as in cool stars, in low-flux as well as in
high-flux sources,  is indicative of problems with the RSRF.}
\item{ From 4.75\,$\mu$m till 4.85\,$\mu$m, a slight increase ($\sim 1$\,\%) is
noticeable in both plots. Once more, an indication for a small problem with the
RSRFs.} 
\item{At the end of band 2A, a quite different behaviour for all the
stars emerges. Further inspection indicates that we see a combination
of problems with the memory-effect correction, synthetic predictions
and, con\-se\-quent\-ly, maybe also with the RSRFs. When we look at Fig.\
\ref{LD_fig6}, we can recognize a same --- increasing --- trend for $\beta$
Peg, $\alpha$ Cet, $\beta$ And and $\alpha$ Tau, the coolest stars in
our sample and so indicating a problem with the synthetic
predictions. This behaviour is however not visible for HD~149447,
whose spectral type is in between $\beta$ And and $\alpha$ Tau. An
analogous discrepancy is visible for $\delta$ Dra with respect to the
stars with almost the same effective temperature. Looking now at Fig.\
\ref{LD_fig7}, we see that HD~149447 and $\delta$ Dra (together with
Vega and $\alpha$ Tuc) have almost the same absolute flux-level in
band 2A and show the same trend (decreasing slope) in this plot,
indicating a problem with the correction for memory-effects, a
statement which is confirmed when we compare the upscan and downscan data.}
\end{itemize}
The majority of discrepancies encountered in bands 2B and 2C are due to
still existing problems with the memory-effect correction and smaller
problems with the RSRFs. This kind of exercise gives us, however, also
an idea on the relative accuracy in band 2: an
accuracy of better than 6\,\% is reached in band 2, which is ---
taking into account the problematic behaviour of the detectors in
band 2 --- a very good result! The new extended model for
memory-effect correction which is now under development by Kester, in
conjunction with a last iteration in the determination of the RSRFs
may even improve this number.

\vspace*{-2ex}
\section{CONFRONTATION WITH THE SEDs OF COHEN} \label{LD_sec:cohen}

\begin{figure*}[t]
\resizebox{.9\textwidth}{!}{\rotatebox{0}{\includegraphics{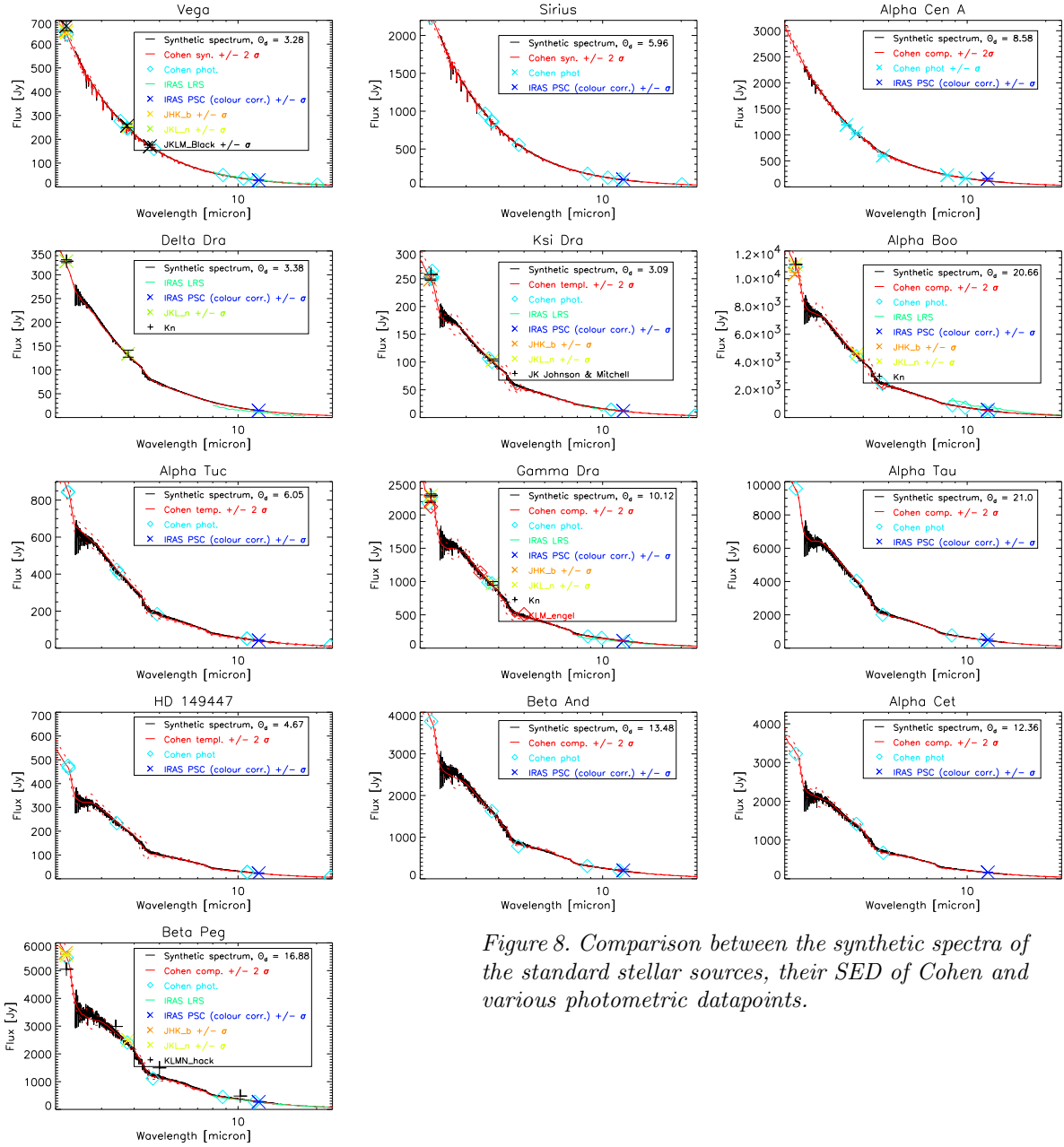}}}\\\\[-10\baselineskip]
\hspace*{.4\textwidth}
\parbox{.45\textwidth}{\caption{\label{LD_fig8}Comparison between the synthetic
spectra of the standard stellar sources, their SED of Cohen and
various photometric datapoints.}}  
\vspace*{5\baselineskip}
\end{figure*}

\begin{figure*}[t]
\resizebox{\textwidth}{!}{\rotatebox{0}{\includegraphics{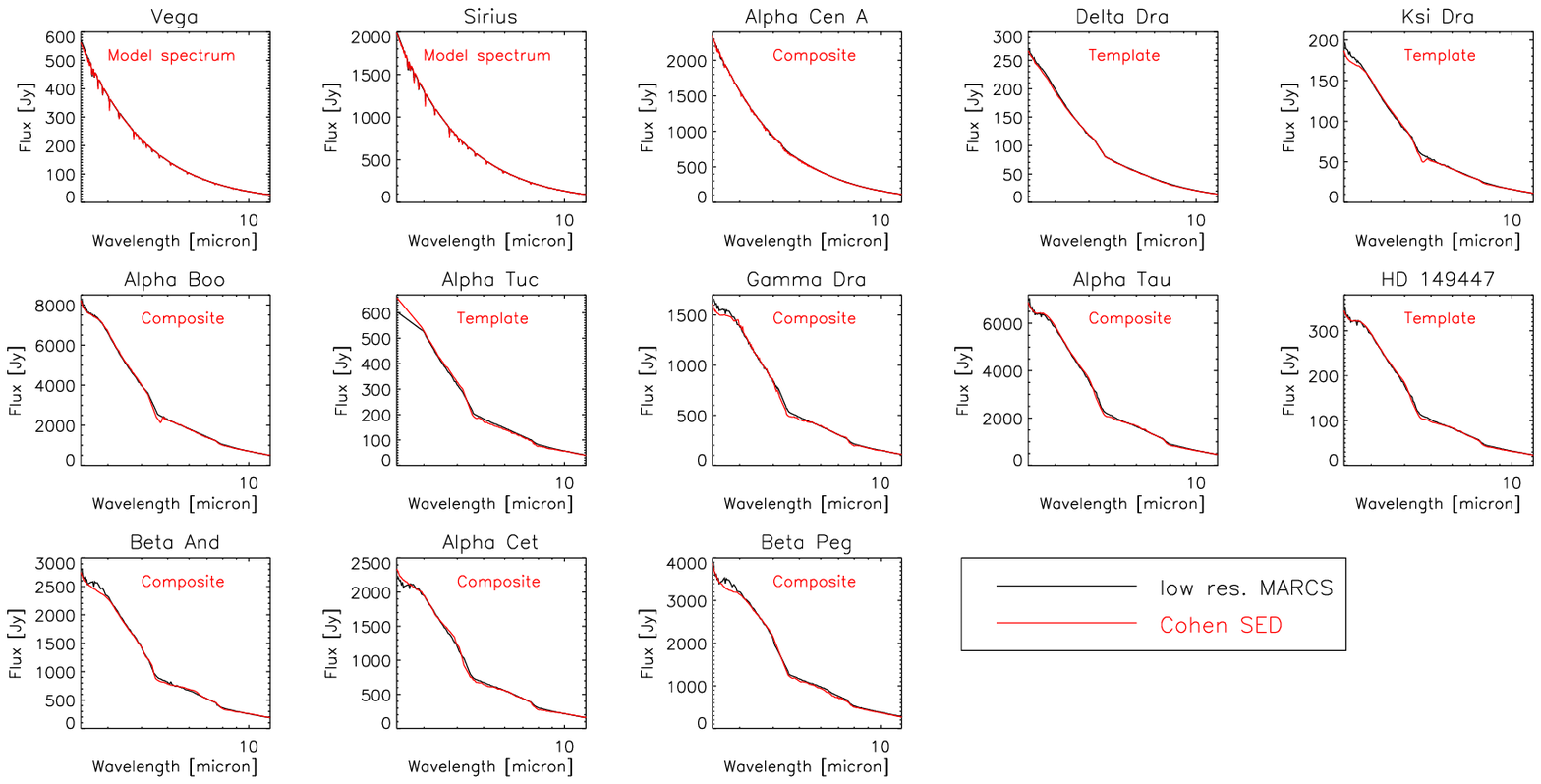}}}
\caption{\label{LD_fig9}Comparison between the SEDs of Cohen (red) for
the standard stellar 
sources and the synthetic spectra (black) rebinned to the same resolution as
the SEDs of Cohen.} 
\end{figure*}

Since the SEDs by Cohen (\cite{LD_cohen1992}, \cite{LD_cohen1995},
\cite{LD_cohen1996}, \cite{LD_witteborn1999})
are used for the calibration of the SWS spectrometers and the
synthetic spectra resulting from this project were used to
improve the calibration in OLP10, a confrontation between the two sets
of data may be instructive. Several stars are in common between the
two samples: 1.\ Vega and Sirius for which Cohen has constructed a
calibrated model spectrum; 2.\ a composite spectrum (i.\ e.\ various
observed spectra have been spliced to each other using photometric
data) is available for $\alpha$ Cen A, $\alpha$ Boo, $\gamma$ Dra,
$\alpha$ Tau, $\beta$ And, $\alpha$ Cet and $\beta$ Peg; 3.\ a
template spectrum (i.\ e.\ a spectrum made by using photometric data of
the star itself and the shape of a `template' star) is built for
$\delta$ Dra (template: $\beta$ Gem: K0~III), $\xi$ Dra (template:
$\alpha$ Boo: K2~IIIp), $\alpha$ Tuc (template: $\alpha$ Hya:
K3~II-III) and HD~149447 (template: $\alpha$ Tau: K5~III). One should
notice that for the composite spectrum of $\gamma$ Dra, Cohen didn't
have any spectroscopic data available in the wavelength range from
1.2\,$\mu$m till 5.5\,$\mu$m. The spectrum of $\alpha$ Tau has
therefore been used to construct the composite spectrum of $\gamma$
Dra in this wavelength range. In order to
ensure independency of the absolute-flux level of the ISO-SWS data, the
angular diameters of the synthetic spectra were determined by using the
photometric data of Cohen, JKLM data from Hammersley's GBPP broad band
photometry (\cite{LD_hammersley2001}), IRAS data and some other
published photometric data cited in the IA\_SED data-base. A comparison
between the (high-resolution) synthetic spectra, SEDs of Cohen and
used photometric data is given in Fig.\ \ref{LD_fig8}. The obtained
angular diameters $\theta_d$ (in mas) for the stellar sources are
mentioned in the legend. To better judge upon the relative agreement
of these two data-sets, the synthetic spectra were rebinned to the
same resolution as the SEDs of Cohen (see Fig.\ \ref{LD_fig9}). The
most remarkable discrepancies between the two spectra arise in the CO
and SiO molecular bands, where the molecular bands are consistently
across our sample stronger
in the composites of Cohen. Cohen has used low-resolution NIR and KAO
data to construct this part of the spectrum. Comparing our synthetic
spectra with OLP8.4 ISO-SWS data (whose calibration is not based on
our results) and the high-resolution Fourier Transform Spectrometer
(FTS) spectrum of $\alpha$ Boo published by \cite*{LD_hinkle1995} (see
Fig.\ 4.5 in \cite{LD_decin2000th}), 
we did however see that the strongest, low-excitation, CO lines were
always predicted as being somewhat too {\it strong} (a few percent at
a resolution of 50000 for the FTS spectrum), probably caused by a
problem with the 
temperature distribution in the outermost layers of the theoretical
model (\cite{LD_decin2001_IV}). Moreover, a comparison between
our high-resolution synthetic spectrum of $\alpha$ Boo with its
FTS spectrum shows also a good agreement for the high-excitation CO
$\Delta v = 1$ lines, the low-excitation lines being predicted as too
weak. The few percent disagreement between FTS and synthetic
spectrum will however never yield the kind of disagreement one
sees between the low-resolution synthetic spectra and the SED
data. The CO $\Delta v = 2$ discrepancy visible in $\gamma$ Dra results
from using the 
$\alpha$ Tau observational KAO data. Since $\alpha$ Tau and $\gamma$
Dra have a different set of stellar parameters, with $\alpha$ Tau
having a significant larger amount of carbon, the SED of Cohen for
$\gamma$ Dra displays too strong a CO feature. One should also be
heedful of this remark when judging upon the quality of the template
spectra for $\delta$ Dra, $\xi$ Dra, $\alpha$ Tuc and HD~149447. 

As conclusion, we may say that the SED spectra of Cohen are excellent
(and consistent) for the absolute calibration of an instrument, but
that attention should be paid when using them for relative flux calibration.

\vspace*{-1ex}
\section{CONCLUSIONS: LESSONS LEARNED} \label{LD_sec:concl}

The purpose of this study was to investigate standard stellar sources
observed with ISO-SWS in order to improve the calibration of the SWS
spectrometers and to elaborate on new theoretical developments for the
modelling of these stars. In spite of the moderate resolution of
ISO-SWS, the stellar parameters for the cool giants could be pinned
down very accurately from their SWS data. This set of consistent IR
synthetic spectra could then be used as input for the OLP10
calibration tools. We could demonstrate that the relative accuracy has
by now reached the 2\,\% level in band 1 and the 6\,\% level in band 2
for (high-flux) sources observed with ISO-SWS. A comparison with the
SEDs of Cohen shows us that this set of 16 synthetic spectra are a
necessary completion for an instrumental calibration refined at the
highest possible level. This kind of analysis, which has been proven
to be very adequate, will be applied to bands 3 and 4 and has opened
also new channels of research, not only for other instruments on board
ISO, like ISO-CAM, but also for the determination of the sensitivity
of future telescopes and satellites, like MIDI and Herschel.

\phantom{\object{$\alpha$ Lyr},\object{$\alpha$ CMa},\object{$\beta$
Leo},\object{$\alpha$ Car},\object{$\alpha$ Cen A},\object{$\delta$
Dra},\object{$\alpha$ Boo},\object{$\alpha$ Tuc},\object{$\beta$
UMi},\object{$\gamma$ Dra},\object{$\alpha$ Tau},\object{HD~149447},
\object{$\beta$ And},\object{$\alpha$ Cet},\object{$\beta$ Peg}}

\vspace*{-1ex}
\begin{acknowledgements}
LD would like to thank C.\ Waelkens, B.\ Vandenbussche, K.\ Eriksson,
B.\ Gustafsson, B.\ Plez and A.\ J.\ Sauval for many fruitful discussions.

\end{acknowledgements}

\end{document}